\documentclass[11pt,a4paper]{article}
\usepackage{amsmath,amssymb,mathrsfs}
\usepackage[dvips]{graphicx}
\usepackage{epsfig}

\newcommand{\ms}[1]{\mathscr{#1}}
\newcommand{\mc}[1]{\mathcal{#1}}

\def\un#1{\relax\ifmmode\@@underline#1\else
        $\@@underline{\hbox{#1}}$\relax\fi}

\let\du=\du                     

\def\f{\phi}

\def\F{\Phi}

\def\bo{{\raise-.3ex\hbox{\large$\Box$}}}               
\def\TH{{\raise.2ex\hbox{$\displaystyle \bigodot$}\mskip-4.7mu \llap H \;}}
\def\face{{\raise.2ex\hbox{$\displaystyle \bigodot$}\mskip-2.2mu \llap {$\ddot
        \smile$}}}                                      

\def\Bar#1{\overline{#1}}                       
\def\leftrightarrowfill{$\mathsurround=0pt \mathord\leftarrow \mkern-6mu
        \cleaders\hbox{$\mkern-2mu \mathord- \mkern-2mu$}\hfill
        \mkern-6mu \mathord\rightarrow$}
\def\dvec#1{\vbox{\ialign{##\crcr
        \leftrightarrowfill\crcr\noalign{\kern-1pt\nointerlineskip}
        $\hfil\displaystyle{#1}\hfil$\crcr}}}           

\def\sfrac#1#2{{\vphantom1\smash{\lower.5ex\hbox{\small$#1$}}\over
        \vphantom1\smash{\raise.4ex\hbox{\small$#2$}}}} 
\def\bfrac#1#2{{\vphantom1\smash{\lower.5ex\hbox{$#1$}}\over
        \vphantom1\smash{\raise.3ex\hbox{$#2$}}}}       
\def\afrac#1#2{{\vphantom1\smash{\lower.5ex\hbox{$#1$}}\over#2}}    

\def\[{\lfloor{\hskip 0.35pt}\!\!\!\lceil}
\def\]{\rfloor{\hskip 0.35pt}\!\!\!\rceil}

\def\du#1#2{_{#1}{}^{#2}}

\def\ha{{\fracmm12}}

\def\un{\underline}
\def\fracmm#1#2{{{#1}\over{#2}}}

\def\low#1{{\raise -3pt\hbox{${\hskip 0.75pt}\!_{#1}$}}}

\newskip\humongous \humongous=0pt plus 1000pt minus 1000pt

\newif\ifdtup

\newcommand{\be}{\begin{equation}}
\newcommand{\ee}{\end{equation}}
\newcommand{\nbe}{\begin{equation*}}
\newcommand{\nee}{\end{equation*}}

\newcommand{\lb}{\label}

\def\lessim{\lower0.6ex\hbox{$\,$\vbox{\offinterlineskip\hbox{$<$}\vskip1pt\hbox{$\sim$}}$\,$}}
\def\grtsim{\lower0.6ex\hbox{$\,$\vbox{\offinterlineskip\hbox{$>$}\vskip1pt\hbox{$\sim$}}$\,$}}

\newcommand{\vev}[1]{\left\langle #1\right\rangle}

\begin{document}

\begin{titlepage}

\begin{center}

June 2014 \hfill IPMU14-0128\\
\hfill UT-14-27

\noindent
\vskip3.0cm
{\Large \bf 

Inflation in Supergravity with a Single Chiral Superfield
}

\vglue.3in

{\large
Sergei V. Ketov~${}^{a,b,c}$ and Takahiro Terada~${}^{d}$ 
}

\vglue.1in

{\em
${}^a$~Department of Physics, Tokyo Metropolitan University \\
Minami-ohsawa 1-1, Hachioji-shi, Tokyo 192-0397, Japan \\
${}^b$~Kavli Institute for the Physics and Mathematics of the Universe (IPMU)
\\The University of Tokyo, Chiba 277-8568, Japan \\
${}^c$~Institute of Physics and Technology, Tomsk Polytechnic University\\
30 Lenin Ave., Tomsk 634050, Russian Federation \\
${}^d$~Department of Physics, The University of Tokyo, Tokyo 113-0033, Japan
}

\vglue.1in
ketov@tmu.ac.jp, takahiro@hep-th.phys.s.u-tokyo.ac.jp

\end{center}

\vglue.3in

\begin{abstract}
\noindent We propose new supergravity models describing chaotic Linde- and Starobinsky-like inflation in terms of a 
\textit{single} chiral superfield. The key ideas to obtain a positive vacuum energy during large field inflation are (i)  stabilization of the real or imaginary partner of the inflaton by modifying a K\"ahler potential, and (ii) use of the crossing terms in the scalar potential originating from a polynomial superpotential. Our inflationary models are constructed by starting from the minimal K\"{a}hler potential with a shift symmetry, and are extended to the no-scale case. Our methods can be applied to more general inflationary models in supergravity with only one chiral superfield.

\end{abstract}

\end{titlepage}


\section{Introduction}\label{sec:Intro}

In the single-field inflationary models the availability of arbitrary choice of the inflaton scalar potential $V(\f)$
allows one to theoretically describe {\it any} value of the Cosmic Microwave Background (CMB) observables $(n_{\text{s}},r)$.  It is one of the reasons
for popularity of the much more restrictive {\it one}-parameter single-field inflationary models with the quadratic ($V=\ha m^2 \f^2$), quartic ($V= \frac{1}{4!} \lambda \f^4$), or the Starobinsky $(V(\f) = \frac{3}{4} M^2( 1- e^{-\sqrt{2/3}
\f })^2$) scalar potentials of the canonically normalized inflaton scalar field $\f$.

For instance, the Starobinsky model \cite{star1} has only one mass parameter $M$ that is fixed by the observational (CMB) data as  $M=(3.0 \times10^{-6})(50/N_e)$ where $N_e$ is the e-foldings number.
The predictions of the Starobinsky model for the spectral index  $n_{\text{s}}\approx 1-2/N_e\approx 0.96$ (for $N_e =50$), the tensor-to-scalar ratio $r\approx 12/N^2_e\approx0.0048 $ and low non-Gaussianity are in agreement with the WMAP and Planck data ($r<0.13$ and $r<0.11$, respectively, at 95\% CL) \cite{Hinshaw:2012aka, planck2}, but are in disagreement with the BICEP2 measurements ($r=0.2^{+0.07}_{-0.05}$, or $r=0.16^{+0.06}_{-0.05}$ when dust subtracted) \cite{bicep2}. The competitive model of chaotic inflation with a quadratic scalar potential, proposed by Linde \cite{linde}, predicts $r\approx 8/N_e = 0.16\left(50/N_e \right)$ in good agreement with the BICEP2 data, though in apparent disagreement with the Planck data (when running of the spectral index is ignored).

It is desirable to realize those single-field inflationary models (and their known extensions, in order to accommodate both the Planck and BICEP2 data) in \textit{supergravity} because supersymmetry (SUSY) is one of the leading candidates for new physics beyond the Standard Model.
 At the same time, we would like such model building to be {\it minimalistic} in the sense that a number of fields and parameters should be limited or constrained as much as possible. 

It is {\it not} straightforward to extend the single-field inflationary models to $4D$, $\mc{N}=1$ supergravity. In the context of the old-minimal supergravity, it requires a complexification of inflaton and its embedding into a scalar supermultiplet whose generic action is parameterized by a K\"ahler potential $K(\F,\Bar{\F})$ and a superpotential $W(\F)$.  The $F$-type scalar potential of supergravity in the Einstein frame is \cite{crem}
\begin{align} \lb{cremf}
V=e^{K}\left( K^{\bar{\Phi} \Phi} \left| W_{\Phi} + K_{\Phi} W \right|^{2} -3 \left| W \right|^{2} \right)
\end{align}
where the subscripts denote the differentiation, and the superscripts stand for the inverse matrix.
There are two obstacles to realize inflation with this potential. First, the exponential factor generically prevents a flat inflationary potential. Second, the scalar potential in terms of a single chiral superfield tends to become negative in a large field region (Sec.~\ref{sec:chaotic}). 

A way to overcome those obstacles was proposed in Refs.~\cite{kawa,kl1,klr} by the use of {\it two} chiral superfields $\F$ and $S$ with a \textit{shift-symmetric} K\"ahler potential and the following superpotential:
\be\lb{kah1}
K = K(\F+\Bar{\F},\Bar{S}S) ,
\qquad 
W = Sf(\F)~,
\ee
respectively.

This choice leads to the very simple scalar potential
\be\lb{pot1}
V=|f(\F)|^2
\ee
in terms of the leading (and canonically normalized) complex scalar field component $\F$~\footnote{
The leading field component of a superfield and the superfield itself are denoted by the same letters.}, provided that the K\"ahler potential is quadratic in $(\F+\Bar{\F})$ {\it and} the superfield $S$ is stabilized at $S=0$.~\footnote{It is not difficult to properly choose the $\Bar{S}S$-dependence of the  K\"ahler potential   (\ref{kah1}) for that purpose.} The supergravity extension given above describes a large (parameterized by the
function $f$) class of non-negative scalar potentials suitable for inflation. For instance, the quadratic single-field scalar potential is easily generated by choosing $f(\F)=m\F/\sqrt{2}$, with Im$\,\F$ playing the role of inflaton along
the inflationary trajectory $S={\rm Re}\,\F=0$. Similarly, the Starobinsky inflation can be described by identifying
inflaton (scalaron) with Re$\,\F$ under the choice $f(\F) =\frac{\sqrt{3}}{2} M (1-e^{-\sqrt{2/3}\F })$ after stabilization of the other fields at  $S={\rm Im}\,\F=0$.

The {\it minimal} K\"{a}hler potential, having the shift symmetry and used in Refs.~\cite{kawa,kl1,klr}, 
\begin{align}
K=\frac{1}{2}\left( \Phi + \bar{\Phi} \right)^{2}~,
\end{align}
leads to free kinetic terms.  Another ({\it no-scale}) K\"{a}hler potential with the shift symmetry is well motivated by (perturbative) superstring compactification \cite{witten} and supersymmetric particle phenomenology beyond the Standard Model \cite{noscale,nsph},
\be \lb{nosca}
K =- 3 \ln  \left( \F + \bar{ \F} -  \frac{1}{3} \bar{S}S  \right) ~,
\ee
where the two complex superfields $(\F,S)$ parameterize the non-compact homogeneous space $SU(2,1)/U(2)$. In the context of superstrings, $\F$ has physical interpretation as a K\"ahler (or compactified volume) modulus. The no-scale supergravity in terms of two chiral superfields $(\F,S)$ with the K\"ahler potential (\ref{nosca}) and a superpotential $W(\F,S)$ was used to embed the Starobinsky inflation in Refs.~\cite{nos0,nos1} and to embed the quadratic (Linde) inflation in Ref.~\cite{nos2}.  

The superfield $S$ above is introduced to obtain the suitable potential of $\Phi$, so that the $S$ plays the auxiliary role.
Our purpose in this Letter is to get rid of the superfield $S$. There are models with a single chiral (or linear) superfield and a real (vector) superfield~\cite{gre, fklp1, Farakos:2014gba}, as well as models with a dynamical complex scalar field based on the non-linear realization of supergravity that requires additional constrained or auxiliary superfields~\cite{Antoniadis:2014oya}. Our construction does not require non-linear constraints, auxiliary superfields, or vector superfields.

As a matter of fact, there are the negative statements against such construction in the literature. As regards a quadratic inflation in the $SU(1,1)/U(1)$ no-scale supergravity, it was noticed in Ref.~\cite{nos2}  under 
Eq.~(24) that ``there are no polynomial forms of $W(\F)$ that lead to a quadratic potential for a canonically normalized field, and we are led to consider $N\geq 2$ models with additional matter fields'' (like $S$).
Also, an extension of the original $(R+R^2)$  Starobinsky model to higher-derivative supergravity does require the second superfield $S$ --- see, e.g., Refs.~\cite{fkpro,kt2} for details.
It has also been shown that it is impossible to embed the Starobinsky model into standard supergravity with the $SU(1,1)/U(1)$ no-scale K\"{a}hler potential when using a \textit{monomial} superpotential~\cite{nos1}.

 However, it is still unclear (a) whether the extra (matter) superfield $S$ is truly necessary in order to embed a quadratic inflation into 
  supergravity, and (b) whether the Starobinsky scalar potential can be embedded into standard supergravity with a single chiral matter superfield with more general K\"{a}hler- and super-potentials.  Needless to say, both issues are highly actual in light of the BICEP2 and Planck data.
 In this Letter we are going to show by explicit construction that it is possible to realize supergravity models of the quadratic (Linde) inflation and of the Starobinsky-like inflation  by using a  K\"ahler potential $K(\F,\bar{\F})$ and a superpotential $W(\F)$, in terms of a {\it single} chiral superfield $\F$. 

 Our paper is organized as follows. In Sec.~\ref{sec:chaotic} we formulate a new class of the supergravity models with a generic Wess-Zumino superpotential in terms of a single chiral superfield $\F$, and demonstrate how to get a quadratic potential of the imaginary part of the $\F$-scalar (identified with axion, similarly to the
natural inflation \cite{free}).  We also explain how to stabilize the second scalar given by the real part of $\F$. In Sec.~\ref{sec:Starobinsky} we apply the same strategy for realizing the Starobinsky-like inflation in supergravity. Our conclusion is Sec.~\ref{sec:conclusion}.

\section{Chaotic inflation with a single chiral superfield}\label{sec:chaotic}
The scalar potential of supergravity \eqref{cremf}
is a sum of the positively definite (first) term and the negatively definite (second) term. To obtain a positive potential in the large field region, the first term should dominate over the second one. Due to a presence of the exponential factor, having a shift symmetry is the key to realize a flat potential in supergravity.

Let us begin with a minimal K\"{a}hler potential having a shift symmetry,
\begin{align}
K=\frac{1}{2} \left( \Phi + \bar{\Phi} \right)^{2},
\end{align}
and a monomial superpotential
\begin{align} 
W=c_n \Phi^n  \quad ({\rm no~ sum})~. \label{monomial}
\end{align}
Then the scalar potential is given by
\begin{align} \lb{scalone}
V=e^{\frac{1}{2} \left( \Phi + \bar{\Phi} \right)^{2}} |c_n |^2 |\Phi |^{2n-2} \left( \left| n+ \left( \Phi + \bar{\Phi} \right) \Phi \right| ^2 -3 | \Phi |^2  \right).
\end{align}
If inflation is driven by the imaginary part of $\Phi$, the real part of $\Phi$ vanishes because of the $Z_2$ symmetry. But then the factor $K_{\Phi}=\Phi+ \bar{\Phi}$ drops out, and the negative term dominates the potential. Note that the shift symmetry of the K\"{a}hler potential eliminates the inflaton (imaginary part) from the K\"{a}hler potential and its derivatives, so that the large field behavior of the inflaton gets worse. Even if one takes a polynomial superpotential breaking the $Z_2$ symmetry, as long as the value of real part of $\Phi$ is of the order of the inflationary scale, which is much lower than the reduced Planck scale, the situation does not significantly change. To improve the situation, there are two possibilities: either to enhance the positively definite term, or to suppress the negatively definite term. Most of the literature uses the second opportunity. In particular,
the sGoldstino field $S$, whose value is fixed to zero, is introduced to eliminate the negatively definite term.

Here we employ the alternative (first) option. We stabilize the real part of $\Phi$ by a higher dimensional operator in the K\"{a}hler potential,
\begin{align}
K=\frac{1}{2} \left( \Phi + \bar{\Phi} \right)^{2} - \zeta \left( \Phi + \bar{\Phi} - 2\Phi_0 \right)^{4}, \label{minimalKstabilized}
\end{align}
so that the factor $|K_\Phi |$ becomes larger than $\sqrt{3}$.
We assume a sufficiently large value of $\zeta$ so that the expectation value of the real part of $\Phi$ is well approximated by $\Phi_0$. The inflaton scalar potential reads
\begin{align}
V=e^{2\Phi_{0}^{2}} |c_n|^{2} \left(\Phi_0 ^2+ \left(\text{Im}\Phi \right)^2 \right)^{n-1} \left[  n^2+ \Phi_0^2 +4\Phi_0^4 +\left( 4\Phi_0^2 -3 \right)(\text{Im} \Phi )^2 \right]~.
\end{align}
With $n=1$ we obtain a quadratic potential of the inflaton $\text{Im} \Phi$.
The cosmological constant can be adjusted to zero (Minkowski vacuum) by adding a constant $c_0$
 to the superpotential without affecting the inflationary potential,
\begin{align}
W=c_0 + c_1 \Phi,
\end{align}
where $\tilde{c}_0=c_0/c_1$ is a solution to $(4\Phi_0^2 -3)\tilde{c}_0^2+(8\Phi_0^3-2\Phi_0 )\tilde{c}_0 + 4\Phi_0^4+\Phi_0^2+1=0$.
It yields
\begin{align}
V=e^{2\Phi_{0}^{2}} |c_1|^{2}  \left( 4\Phi_0^2 -3 \right)(\text{Im} \Phi )^2 .
\end{align}
The real constant $\Phi_0$ must be larger than $\sqrt{3}/2$, avoiding a tachyonic inflaton mass.

The same strategy can be applied to the no-scale K\"{a}hler potential.
Let us consider the no-scale K\"{a}hler potential with a generic superpotential,
\be \lb{noks}
K=-3\ln \left( \frac{\Phi+\bar{\Phi}}{3} \right)~, \qquad W=W(\Phi)~,
\ee
leading to the following kinetic term and the scalar potential:
\begin{align} \lb{setup_kin}
\mc{L}_{\text{kin}}=&-\frac{3}{(\Phi+\bar{\Phi})^{2}} \partial_{\mu}\bar{\Phi}\partial^{\mu}\Phi=-\frac{3}{4(\text{Re}\Phi)^{2}}\left[ (\partial_{\mu}\text{Re}\Phi)^{2}+(\partial_{\mu}\text{Im}\Phi)^{2} \right], \\
V=&\frac{9}{(\Phi+\bar{\Phi})^{2}}\left[ (\Phi+\bar{\Phi})\left| W_{\Phi} \right|^{2}-3\left( \bar{W}W_{\Phi}+W\bar{W}_{\bar{\Phi}}\right) \right]. \label{setup_pot}
\end{align}
The no-scale property means that $V=0$ for $W=c_3\Phi^3$ with any value of the coupling constant $c_3$.

It is worth mentioning that the standard matter-coupled supergravity defined by Eq.~(\ref{noks}) is dual (equivalent)
to the so-called $F(\mc{R})$ higher-derivative supergravity defined by the chiral superspace action \cite{gket}
\begin{align} \label{FR}
S_{F}= \left[ \int \mathrm{d}^{4}x\mathrm{d}^{2}\Theta 2 \ms{E} F(\mc{R})+\text{H.c.}\right]~,
\end{align}
whose holomorphic function $F$ of the chiral scalar curvature superfield $\mc{R}$ is related to the chiral
superpotential $W(\Phi)$ via the Legendre transform \cite{ks1,ks2,krev}.

In the case of a generic monomial superpotential \eqref{monomial}, the scalar potential reads
\begin{align}
V=\frac{9n(n-3)}{\Phi+\bar{\Phi}}|c_n|^2 |\Phi|^{2n-2} \label{no-scale_mono_potential}
\end{align}
and is positive only when $n>3$ or $n<0$.   To gain more insight, let us take a polynomial superpotential 
having two terms,
\begin{align}
W=c_{m} \Phi^{m} + c_{n} \Phi^{n}, \label{2term_superpotential}
\end{align}
and natural powers $n>m$. Then the scalar potential is
\begin{align}
V=& \frac{9}{\left( \Phi +\bar{\Phi } \right)^{2}} \left\{ \left(\Phi + \bar{\Phi} \right) \left[ \left(n^2 - 3n \right) \left| c_{n} \right|^{2}\left| \Phi \right|^{2n-2} + \left(m^2 - 3m \right) \left| c_{m} \right|^{2}\left| \Phi \right|^{2m-2}+\right.
  \right. \nonumber \\
& \quad    \left. \left. +2nm \text{Re} \left( c_{n}\bar{c_{m}} \bar{\Phi}^{m-1}\Phi^{n-1} \right) \right] - 6m \text{Re}\left( \bar{c_{n}}c_{m}\bar{\Phi}^{n}\Phi^{m-1} \right) -6n \left( \bar{c_{m}}c_{n}\bar{\Phi}^{m}\Phi^{n-1} \right) \right\} .\label{2term_potential}
\end{align}
Since the Re\,$\Phi$ will be eventually stabilized (see below), we can regard the functions of Re$\,\Phi$ 
as the coefficients of the
 polynomial of Im$\,\Phi$ in Eq.~(\ref{2term_potential}). We want a quadratic potential for $\text{Im}\Phi$. The largest power of $\text{Im}\Phi$ in Eq.~\eqref{2term_potential} is $2n-2$, while $m+n-1$ is of the same value provided that $m=n-1$.  We want these to be quadratic. When demanding $2n-2=2$, we need $m+n-1=2$, {\textit i.e.} $n=2$ and $m=1$, in order to get a positive overall coefficient.
Alternatively, one may take $n=3$ and generate a quadratic term by using the cross terms of the $(m+n-1)$th power.  In this case $m=0$.

Next, let us take the most general renormalizable (Wess-Zumino) superpotential,
\begin{align} \lb{potan}
W=c_0 + c_1 \Phi + \frac{1}{2} c_2 \Phi^{2} + \frac{1}{3} c_3 \Phi^{3}~,
\end{align}
with arbitrary (complex) coupling constants $(c_0,c_1,c_2,c_3)$. It yields the scalar potential
\begin{align}\lb{gspot}
V=&-\frac{27}{2(\text{Re}\Phi)^{2}} \left\{ \text{Re}(\bar{c_0} c_1 )+\left( \text{Re}(\bar{c_0} c_2)+\frac{2}{3}|c_1|^{2}\right)\text{Re}\Phi +\left( \text{Re}(\bar{c_0} c_3)+\frac{5}{6}\text{Re}(\bar{c_1}c_2) \right)(\text{Re}\Phi)^{2} \right. \nonumber \\
&  +\left( \frac{2}{3}\text{Re}(\bar{c_1}c_3)+\frac{1}{6}\left| c_2\right|^{2}\right)(\text{Re}\Phi)^{3}+\frac{1}{6}\text{Re}(\bar{c_2}c_3)(\text{Re}\Phi)^{4} \nonumber \\
&   +\left[ -\text{Im}(\bar{c_0}c_2)+\left( \frac{1}{3}\text{Im}(\bar{c_2}c_1)-2\text{Im}(\bar{c_0}c_3) \right)\text{Re}\Phi-\frac{2}{3}\text{Im}(\bar{c_1}c_3)(\text{Re}\Phi)^{2} \right] \text{Im}\Phi  \nonumber \\
&  +\left[ \frac{1}{2}\text{Re}(\bar{c_2}c_1)-\text{Re}(\bar{c_0}c_3)+\left(  \frac{2}{3}\text{Re}(\bar{c_1}c_3)+\frac{1}{6}\left|c_2\right|^{2} \right) \text{Re}\Phi +
\frac{1}{3}\text{Re}(\bar{c_2}c_3)(\text{Re}\Phi)^{2} \right](\text{Im}\Phi)^{2}  \nonumber \\
&  \left.  -\frac{2}{3}\text{Im}(\bar{c_1}c_3)(\text{Im}\Phi)^{3}+\frac{1}{6}\text{Re}(\bar{c_2}c_3)(\text{Im}\Phi)^{4} \right\},
\end{align}
having many mixing terms between the couplings of Eq.~(\ref{potan}) due to
the nonlinearity of the equation (\ref{setup_pot}) with respect to $W$.

Then, after a stabilization of the real part of  $\Phi$, the canonically normalized imaginary part, $\Phi_{\text{I}}$, is given by
\begin{align} \lb{canon1}
\text{Im}\Phi=&\sqrt{\frac{2}{3}} \vev{\text{Re}\Phi}\Phi_{\text{I}},
\end{align}
where the bracket denotes the vacuum expectation value. For instance, setting $c_3=0$ 
eliminates  
both cubic and quartic terms with respect to Im\,$\Phi$, so that the scalar potential in terms of the canonically normalized field $\Phi_{\text{I}}$ is given by 
\begin{align} \lb{canonsp1}
V=&\frac{27}{2}\left[ -\text{Re}(\bar{c_0}c_1)\vev{\text{Re}\Phi}^{-2}-\left( \text{Re}(\bar{c_0}c_2)+\frac{2}{3}\left|c_1\right|^{2} \right) \vev{\text{Re}\Phi}^{-1}-\frac{5}{6}\text{Re}(\bar{c_1}c_2)-\frac{1}{6}\left|c_2\right|^{2}\vev{\text{Re}\Phi} \right] \nonumber \\
&+9\sqrt{\frac{3}{2}}\left( \text{Im}(\bar{c_0}c_2)\vev{\text{Re}\Phi}^{-1}-\frac{1}{3}\text{Im}(\bar{c_2}c_1) \right)\Phi_{\text{I}} +9\left( -\frac{1}{2}\text{Re}(\bar{c_2}c_1)-\frac{1}{6}\left|c_2\right|^{2}\vev{\text{Re}\Phi} \right) \Phi_{\text{I}}^{2}~.
\end{align}
In particular, the mass squared of the inflaton $\Phi_{\text{I}}$ is 
\begin{align} \lb{infmass1}
m_{\Phi_{\text{I}}}^{2}=-9\text{Re}(\bar{c_2}c_1)-3\left|c_2\right|^{2}\vev{\text{Re}\Phi}~.
\end{align}
It is positive for some $c_1$ and $c_2$, and should be matched with the amplitude of curvature perturbations, $\mc{P}_{\mc{R}}(k)=(2.196^{+0.051}_{-0.060})\times 10^{-9} (k/k_0 )^{n_{\text{s}} -1}$ with pivot scale $k_0=0.05 \text{Mpc}^{-1}$~\cite{planck2}, 
so that $m_{\Phi_{\text{I}}}=1.8 \times 10^{13} \text{GeV}$.
The cosmological constant can be adjusted to zero by the choice of $c_0$.
It is also possible to preserve SUSY at the vacuum by tuning of parameters, $c_1=-c_2 \Phi_0$ and $c_0=\frac{1}{2}c_2 \Phi_0^2$, \textit{i.e.} $W=\frac{1}{2}c_2 ( \Phi - \Phi_0 )^2$.

Of course, the whole approach makes sense only when the real part of $\Phi$ is stabilized by modifying the 
K\"ahler potential of Eq.~(\ref{noks}) by some extra terms breaking its no-scale structure.  A particularly simple example of such modification was proposed the long time ago in Ref.~\cite{estab} (and was used more recently in  Refs.~\cite{nos2,genf} for the stabilization purposes with two chiral superfields),
\begin{align} \lb{stab4}
K=-3\ln \left[ \frac{\Phi+\bar{\Phi}+\zeta \left( \Phi+\bar{\Phi}-2\Phi_{0} \right)^{4}}{3}\right]~,
\end{align}
with two real parameters $\zeta$ and $\Phi_{0}$.

It is straightforward to compute the scalar potential with the modified K\"ahler potential. We find
\begin{align} \lb{modpot}
V & = \frac{9}{\left[\Phi+\bar{\Phi}+\zeta (\Phi+\bar{\Phi}-2\Phi_0)^4 \right]^2}\times\nonumber \\
& \frac{1}{  1-4\zeta (\Phi+\bar{\Phi}-2\Phi_0 )^{3}+4\zeta ^2 (\Phi+\bar{\Phi}-2\Phi_0 )^{6}-24\zeta \Phi_0 \left( \Phi+\bar{\Phi}-2\Phi_{0} \right)^{2} } \nonumber \\
&\times  \left[ \left( \Phi+\bar{\Phi}+\zeta (\Phi+\bar{\Phi}-2\Phi_0)^4 \right) \left|W_{\Phi}\right|^{2}- 3\left( 1+4\zeta (\Phi+\bar{\Phi})^{3}\right) \left( \bar{W}W_{\Phi}+W\bar{W}_{\bar{\Phi}}\right)  \right.\nonumber \\
& \left. \phantom{ (\Phi+\bar{\Phi}+\zeta (\Phi+\bar{\Phi}-2\Phi_0)^n \left|W_{\Phi}\right|^{2}}   +108\zeta (\Phi+\bar{\Phi}-2\Phi_0 )^{2}|W|^{2} \right]~.
\end{align}
The real part of $\Phi$ is stabilized by a SUSY breaking mass, 
\be \lb{susybm}
m_{\text{Re}\Phi}^2 \simeq -\left( K^{\bar{\Phi}\Phi} \right)^3 e^{K}K_{\Phi \, \bar{\Phi} \, \Phi \, \bar{\Phi}}\left| W_\Phi + K_\Phi W \right|^2~.
\ee 
Taking a large $\zeta$, it becomes  
$m_{\text{Re}\Phi}^{2} \simeq \left( 648 \zeta - 81/4 \Phi_0^3 \right)|W|^2$ at the vacuum.
It is larger during inflation.

As in the minimal K\"{a}hler case, the factor $(\Phi + \bar{\Phi} -2\Phi_0 )$ is suppressed due to the stabilization effects itself. Therefore, many extra contributions, including those in the canonical normalization, are small compared to the original contribution above, and merely perturb the coefficients of the polynomial in $\Phi_{\text{I}}$.  
The only non-trivial change~\footnote{Strictly speaking, the above corrections may be important for the $n=3$ and $n=0$ terms because of some extra cancellations (cf.~Eq.~\eqref{no-scale_mono_potential}), but they are irrelevant in our example~\eqref{canonsp1}.
} in the potential of $\Phi_{\text{I}}$ comes from the term proportional to $|W|^{2}$ (it represents the gravitational corrections).  In contrast to the case of \eqref{minimalKstabilized}, that term arises because the stabilization term makes a cancellation of the terms characteristic to the no-scale type models {\it incomplete}. The extra term should be small enough, when compared to the original terms, in order  to ensure the quadratically generated chaotic inflationary dynamics. For example, the term responsible for the inflation, 
$\frac{1}{2}m_{\Phi_{\text{I}}}^{2}\Phi_{\text{I}}^{2}$, should be dominant over the quartic term contained in the $|W|^{2}$ term, which implies
\begin{align} \lb{cond2}
-\text{Re}(\bar{c_2}c_1) \gg \frac{3}{2}\zeta (2\text{Re}\Phi-2\Phi_0)^4 |c_2|^{2}\Phi_{\text{I}}^{2}
\end{align}
during inflation. We have assumed here for simplicity that the first term in the mass formula of $\Phi_{\text{I}}$ in Eq.~(\ref{infmass1}) dominates over the second term, which is negatively definite.

\section{Starobinsky-like inflation with a single chiral superfield}\label{sec:Starobinsky}

In this Section we revisit the issue of minimal realization of the Starobinsky-like inflation in supergravity with a single chiral superfield (or $2_{\rm B}+2_{\rm F}$ extra d.o.f. only). The no-go arguments in the literature against it refer to a supergravity extension of $(R+R^2)$ gravity  \cite{fkpro,kt2} in the original (higher-derivative) picture, and a {\it monomial} Ansatz for a superpotential (in the dual picture) \cite{nos1}. 

Since the stabilization \eqref{stab4} leads to breaking of the no-scale structure of the K\"ahler potential, it also breaks its correspondence to the $F(\mc{R})$ higher-derivative supergravity  \eqref{FR}, so that the first no-go
obstruction is dismissed. As regards the second obstruction, in order to achieve a scalar potential which behaves like a constant in the asymptotically large field region, the power $n$ of the superpotential 
 \eqref{monomial} must be $3/2$. Substituting that monomial superpotential into the supergravity scalar potential  \eqref{setup_pot} results in a negative constant (\textit{i.e.} AdS instead of dS) --- see 
 \eqref{no-scale_mono_potential} and Ref.~\cite{nos1}. However, that can be improved by using a {\it polynomial} superpotential.
 
 Let us consider Eqs.~\eqref{2term_superpotential} and \eqref{2term_potential} with $\text{Re} \Phi$ as inflaton,
 and $\text{Im} \Phi$ stabilized as in Eq.~\eqref{stab4} though with $-i (\Phi-\bar{\Phi})$ in the $\zeta$-term. Since  we are interested in the large field behavior of $\text{Re}\Phi$, its largest power is most relevant. Demanding it to be zero yields either $2n-3=0$ that is inappropriate, or $n+m-3=0$, while a larger power due to the 
 $(2n-3)$-term is eliminated by choosing $n=3$ that, in turn, implies $m=0$.

According to those arguments, we should take a superpotential
\begin{align}
W=c_0 + \frac{1}{3}c_3 \Phi^3 \label{03}
\end{align}
as our starting point. After stabilization of the imaginary part ($\text{Im}\Phi=0$), this leads to a constant potential $V=-27\text{Re}(\bar{c_0}c_3)/2$, whose sign can be chosen to be positive.
At this stage, the good news is that we have a positive inflationary energy, and the bad news is that our
scalar potential is just a constant.

\begin{figure}[htbp]
  \begin{center}
    \includegraphics[clip,width=10.0cm]{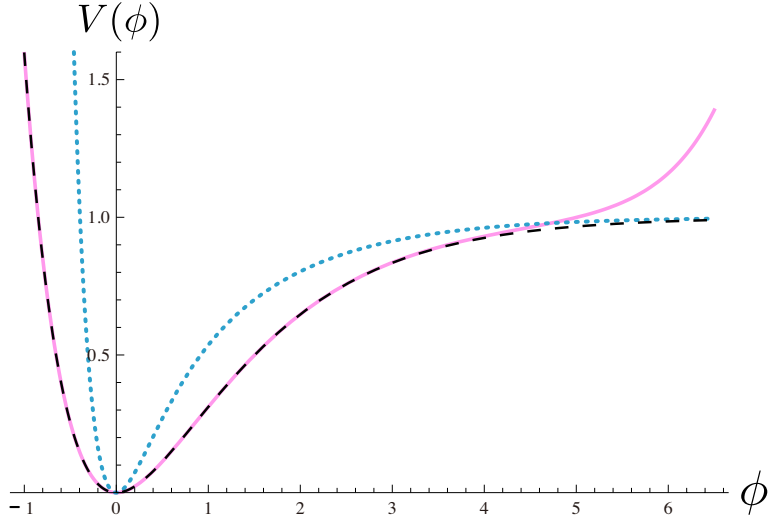}
    \caption{The Starobinsky-like scalar potentials. The pink solid line is of the model~\eqref{03shifted} where the parameter values are taken as $a=1, -27\text{Re}(\bar{c_0}c_3)=1$, and $9a|c_3|^{2}/2=10^{-5}$.  The cyan dotted line is of the model~\eqref{-103}, where the parameters are taken as $a=-b=-c=d=1$. The black dashed line is the original Starobinsky potential. }
    \label{fig:potentials}
  \end{center}
\end{figure}

\begin{figure}[htbp]
  \begin{center}
    \includegraphics[clip,width=10.70cm]{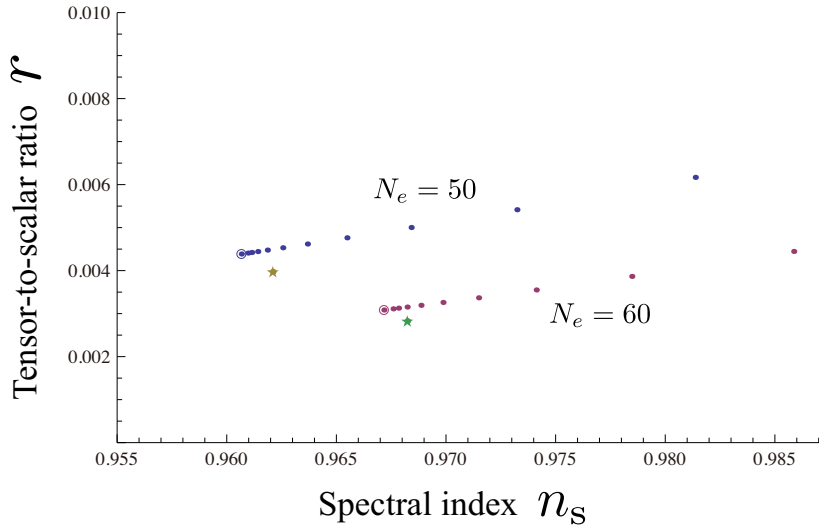}
    \caption{The spectral index and the tensor-to-scalar ratio of the models \eqref{03shifted} and \eqref{-103}.  The prediction of \eqref{03shifted} corresponds to blue points ($N_e =50$) and red points ($N_e =60$).  The very left point (double circle) in each case is the prediction of the Starobinsky model.  The relative coefficient $-c_3 / 3 c_0$ of the correction term is set to $10^{-8}, 10^{-7.8}, 10^{-7.6}, \dots$ from left to right.  We take real parameters and $a=1$.   The prediction of \eqref{-103} corresponds to the yellow ($N_e =50$) and green ($N_e =60$) star, with the same parameters as in Fig.~\ref{fig:potentials}.}
    \label{fig:ns_r}
  \end{center}
\end{figure}

As the next stage, we consider two modifications of the superpotential \eqref{03}, in order to obtain meaningful Starobinsky-like potentials. The first proposal is to introduce a constant shift of the field $\Phi$ in the superpotential as
\begin{align}
W=c_0 + \frac{1}{3}c_3 \left(\Phi-a\right)^3, \label{03shifted}
\end{align}
where the constant $a$ is assumed to be real, so that a cancellation between the contributions of $(\text{Re}\Phi-a)^{2}$ and $(\text{Re}\Phi)^{-2}$ becomes incomplete. After a field redefinition  $\text{Re}\Phi=\exp (\sqrt{2/3}\phi )$, the scalar potential of the canonical field $\phi$ becomes
\begin{align}
V=&-\frac{27}{2}\text{Re}(\bar{c_0}c_3) \left(1-a e^{-\sqrt{2/3}\phi} \right)^{2}+\frac{9}{2}a|c_3|^{2} \left(1-a e^{-\sqrt{2/3}\phi} \right)^{4} e^{2\sqrt{2/3}\phi}.
\end{align}
The first term has the form of the Starobinsky potential, and the second term blows up in the large field region,
with $|\phi |$ of the order $\leq O(10)$ (see the pink solid line in Fig.~\ref{fig:potentials}).  However, if $|c_3|$ is sufficiently small, the latter term can be ignored. In the parameter space, where the second term is small but non-negligible, the predictions of the Starobinsky model for $r$ (Sec.~1) are modified. As an
 illustration, we plot  the prediction of our model in the $(n_{\text{s}}, \, r )$-plane in Fig.~\ref{fig:ns_r}.
We set $a=1$, because increasing $a$ has the same effect as increasing the coefficient of the correction term, and assume the ideal stabilization with $\vev{\text{Im}\Phi}=\Phi_0$.
The e-foldings number $N_e$ is defined between the point where the $(n_{\text{s}}, \, r)$ are evaluated and the point where the slow roll inflation ends, $\epsilon\equiv (V'/V)^{2}/2=1$.

Another (second) way of modification of the Ansatz \eqref{03} is by adding terms of a smaller power to the superpotential, while keeping its large field behavior. As an example, let us consider a
superpotential~\footnote{Allowing singularities and/or meromorphic potentials in field space is the basic feature
of the so-called axion {\it monodromy} inflation in string theory \cite{ura}.}
\begin{align}
W=c_{-1}\Phi^{-1}+c_0 +\frac{1}{3}c_3 \Phi^3. \label{-103}
\end{align}
 After stabilization of the imaginary part, it results in the following scalar potential (the cyan dotted line in Fig.~\ref{fig:potentials}):
\begin{align}
V=& a + b e^{-\sqrt{2/3}\phi}+c e^{-4\sqrt{2/3}\phi}+d e^{-5\sqrt{2/3}\phi},
\end{align}
where $a=-27\text{Re}(\bar{c_0}c_3)/2$, $b=-18\text{Re}(\bar{c_3}c_{-1})$, $c=27\text{Re}(\bar{c_0}c_{-1})/2$, and $d=18|c_{-1}|^{2}$.
SUSY is preserved at the vacuum. 
See Fig.~\ref{fig:ns_r} for the prediction of this model in the $(n_{\text{s}}, \, r )$-plane.

Finally, we point out that the Starobinsky-like potential of the $\text{Im}\Phi$ can also be obtained from a 
non-logarithmic K\"{a}hler potential~\eqref{minimalKstabilized} with a superpotential
\begin{align}
W=& m \left[ b - e^{ia\left( \Phi - \Phi_0 \right)} \right] ,
\end{align}
where we have introduced the mass parameter $m$,  and two real constants, $a$  and $\Phi_0$ (the vacuum expectation value of $\text{Re}\Phi$), and the complex constant $b$. After stabilization of the $\text{Re}\Phi$, the scalar potential of $\text{Im}\Phi$ reads
\begin{align}
V=e^{2\Phi_0^2}|m|^2 \left[ \left( 4\Phi_0^2 -3 \right) \left( \text{Re} b -e^{-a\text{Im}\Phi} \right)^2 +\left( 2\Phi_0 \text{Im}b -a e^{-a \text{Im}\Phi} \right)^2 -3\left( \text{Im}b \right)^2  \right]
\end{align}
and can be a very good approximation to the Starobinsky scalar potential in the large field region. When 
$a=\sqrt{2/3}$ and $\Phi_{0}>\sqrt{3}/2$, there are always two solutions for $\text{Re}b$ and $\text{Im}b$,  realizing the Starobinsky scalar potential.

\section{Conclusion}\label{sec:conclusion}

We demonstrated that it is possible to realize both the Linde (quadratic)- and the Starobinsky-type scalar potentials in
supergravity with a single chiral superfield, without other chiral superfields or any tensor (vector) matter. It requires certain modifications of the K\"ahler potential as well as fine-tuning of the parameters, which are the
common features of all realizations of chaotic inflation in supergravity.  Other stabilization mechanisms are certainly possible, while we used the one of Eqs.~\eqref{minimalKstabilized} and \eqref{stab4} as an example.  The no-scale property can be sacrificed because it does not survive against quantum corrections.  

Some of our models favor the SUSY breaking scale that is much higher than the electroweak scale,
because inflaton breaks SUSY even after inflation. Our realizations lead to super-Planckian excursion of inflaton field \cite{lyth}, though it may not be a problem \cite{dva}. A derivation of the proposed supergravity models from extended supergravity, extra dimensions or string theory is beyond the scope of this paper (see, however, Refs.~\cite{ketn2,ketuni}).

The Linde and Starobinsky models have different shapes of the scalar potential and, hence, lead to different predictions of inflationary observables.  Our method may be used to develop other inflationary models in supergravity with the help of a single chiral superfield. It contributes to the inflationary model building to be consistent with the Planck, BICEP2, and  future data.

\section*{Acknowledgements}

SVK was supported by a Grant-in-Aid of the Japanese Society for Promotion of Science (JSPS) under No.~26400252, and the World Premier International Research Center Initiative (WPI Initiative), MEXT, Japan. TT was supported by a 
Grant-in-Aid for JSPS Fellows, and a Grant-in-Aid of the JSPS under No.~2610619.

\end{document}